\documentclass[11pt]{article}

\usepackage{graphics}  
\usepackage{amsmath}
\usepackage{amssymb}

\newcommand{\D}{\displaystyle}

\usepackage{amsthm}
\theoremstyle{definition}
\newtheorem{theorem}{Theorem}[section]
\newtheorem{lemma}[theorem]{Lemma}

\newtheorem{proposition}[theorem]{Proposition}

\begin{document} 

\title
{\vskip -70pt
\vskip 2cm
{\huge {Gauged vortices in a background}}\\  
\vspace{1mm}
}

\author{
{\Large \bf Nuno M. Rom\~ao}
\thanks{e-mail: {\tt nromao@maths.adelaide.edu.au}}\\[2mm]
{\normalsize \sl School of Pure Mathematics, University of Adelaide}\\
{\normalsize \sl North Terrace, Adelaide SA 5005, Australia} \\[3mm]
}


\maketitle

\begin{abstract}
\noindent
We discuss the statistical mechanics of a gas of gauged vortices 
in the canonical formalism. 
At critical self-coupling, and for 
low temperatures, it has been argued that the configuration space for vortex dynamics in each topological class of the abelian Higgs model
approximately truncates to a finite-dimensional moduli space 
with a K\"ahler structure. For the case where the vortices live on a 2-sphere, 
we explain how localisation formulas on the moduli spaces can be used to 
compute exactly the partition function of the vortex gas interacting with a background potential. The coefficients of this analytic function provide 
geometrical data about the K\"ahler structures, the simplest of which being their symplectic volume (computed previously by Manton using an alternative argument). 
We use the partition function
to deduce simple results on the thermodynamics of
the vortex system; in particular, the average height on 
the sphere is computed and provides an interesting effective picture of the ground state.
\end{abstract}

\vspace{3mm}


\vspace{2mm}


\section{Introduction}\label{secintro}

One of the most challenging aspects in the study of topological solitons
in gauge field theories is to understand their interactions, even at the
classical level. At critical self-coupling, where the solitons exert no net
static forces among themselves, one can typically describe the dynamical interactions at low speed in terms of geodesic flow for certain metrics 
on the moduli spaces of stable field configurations~\cite{ManRk}. Exact 
results 
about these metrics can be obtained in some instances, and they have provided 
detailed information about the classical dynamics of solitons in this 
regime~\cite{ManSut}.
There is now considerable evidence on the beautiful geometrical fact that 
these moduli spaces encode a whole range of physical information about the 
underlying field theories, which goes well beyond the problem of
approximating the slow dynamics that brought them first into mathematical physics.

An example illustrating how physical information can be extracted from the 
geometry of the moduli spaces is provided by the study of the 
statistical mechanics of a gas of vortices in the abelian Higgs model.
Manton obtained
the partition function in the critically 
coupled (noninteracting) regime from the volumes of the moduli spaces~\cite{ManSMV}.
These volumes can be calculated once the K\"ahler
classes of the moduli spaces are known.                         
At first, this calculation was performed for vortices on a sphere, but a similar argument can be used to compute the partition function for vortices living on any compact Riemann surface \cite{ManNasVVM}.

In this paper, the abelian Higgs model is modified by adding to the lagrangian (at critical coupling) an external potential that will probe 
the interactions of the vortices themselves. 
For weak potentials, we expect the effect of this coupling to be well described by the addition of a potential to the moduli space dynamics. In this setting,
we calculate exactly the
partition function for the vortex gas in a background field.
Clearly, the problem that we consider is still less ambitious than the more physically interesting (but also more difficult) statistical mechanics of vortices with Ginzburg--Landau self-interaction, but it does provide a nontrivial extension of the study by Manton.
The vortices will be allowed to live on a sphere with a particular axis singled out, and the potential we shall focus on is natural given this geometry. To obtain the statistical mechanics of the system, we will be
making use of a localisation formula for a circle action on the moduli space
of $N$ vortices. This turns out to be an alternative route to obtain Manton's 
results (which we recover as we switch off the interaction), in particular
his formula for the volume of the moduli space of $N$ vortices on a 
sphere~\cite{ManSMV}.

\section{Gauged vortices and their moduli spaces} \label{secvortices}

Given a Riemann surface $\Sigma$, a gauged vortex is a pair $(d_{A},\Phi)$
consisting of a unitary connection on a hermitian line 
bundle ${\cal L}\rightarrow \Sigma$ and a section of this bundle, 
satisfying the Bogomol'ny\u\i\ (or vortex) equations
\begin{eqnarray}
&\bar{\partial}_{A}\Phi=0,& \label{Bogo1} \\
&B_{A}=\frac{1}{2}*(1-\langle\Phi,\Phi\rangle). \label{Bogo2}&
\end{eqnarray}
Here, $\langle\cdot, \cdot\rangle$ is the hermitian structure
on $\cal L$, and we have
fixed a metric $ds^2_{\Sigma}$ on $\Sigma$ with Hodge star $\ast$; $\bar{\partial}_{A}$ is defined from $d_A=d-iA=:\partial_{A}+\bar{\partial}_{A}$ as usual through the decomposition provided by the complex structure of $\Sigma$, and $B_A=dA$ is the curvature of $d_A$. These equations are invariant under gauge transformations
\[
(d_A,\Phi)\mapsto (d_{A + d\Lambda}, e^{i\Lambda}\Phi),\qquad\Lambda \in 
C^{\infty}(\Sigma; \mathbb{R})\cong \mathfrak{aut}({\cal L},\langle \cdot,\cdot\rangle).
\]
The choice of hermitian structure does not play an important r\^ole in our discussion, so once we fix local trivialisations for ${\cal L}\rightarrow \Sigma$ we use them to pull back the standard hermitian structure of $\mathbb{C}$.

To a configuration $(d_A,\Phi)$ we associate the vortex number
\[
N=\frac{1}{2\pi}\int_{\Sigma}B_{A}.
\]
For the cases of interest, where $\Sigma$ is compact or is effectively
compactified by imposing suitable boundary conditions,
$N\in \mathbb{Z}$ and it corresponds to the degree of ${\cal L}$, a topological invariant.
Since (\ref{Bogo1}) states that $\Phi$ should be a holomorphic section, $N$ 
is also the number of zeroes of $\Phi$, all having positive multiplicity. 
In fact, solutions of (\ref{Bogo1})--(\ref{Bogo2}) are completely 
characterised by the zeroes of $\Phi$, which can be any set of $N$ points on $\Sigma$ (counted with multiplicity)~(cf.~\cite{JafTau}), provided that
\begin{equation} \label{Bradlow}
4\pi N < {\rm Vol}(\Sigma)
\end{equation}
holds~\cite{Nog, Bra}. So the space of solutions to the vortex equations 
modulo gauge equivalence has the structure
\[
\coprod_{N\in\mathbb{N}} {\cal M}_{N},
\]
where each moduli space of $N$-vortices ${\cal M}_{N}$ is the $N$th symmetric power of $\Sigma$, the smooth $2N$-manifold
\[
{\cal M}_{N}={\rm Sym}^{N}(\Sigma):= \Sigma^{N}/{\mathfrak{S}_{N}}.
\]
Complex coordinates on this (complex) manifold are usually referred to as 
moduli. If $z$ is a local coordinate on an open set $U\subset \Sigma$, then 
the natural coordinates $(z_{1},\ldots,z_{N})$ on the cartesian product 
$U^N$, denoting configurations with zeroes of $\Phi$ at each $z=z_r$, form
an appropriate set of moduli on ${\rm Sym}^N(U)-\Delta$, where 
$\Delta\subset {\cal M}_N$ is
the locus on which at least two zeroes become coincident (and thus the
$\mathfrak{S}_N$-action fails to be free). The $z_r$ are interpreted as positions of $N$ individual vortex cores.

The Bogomol'ny\u\i\ equations above appeared for the first time in the study of the abelian Higgs model, a field theory for vortex dynamics defined by the functional
\begin{equation}\label{abelHiggs}
{\cal AH}_{\lambda^2}(D_A,\Phi)=-\frac{1}{4}\| F_{A}\|_{L^2}^{2}+
\frac{1}{2}\|D_{A}\Phi \|_{L^2}^{2} 
-\frac{\lambda^2}{8}\| \langle \Phi,\Phi\rangle -1\|_{L^2}^{2}
\end{equation}
depending on a self-coupling $\lambda^2 \in \mathbb{R}^{+}$. 
The fields $A$ and $\Phi$ here depend on a time parameter $t\in \mathbb{R}$,
and the $L^2$ norms 
are taken with respect to the metric $dt^2-ds^2_{\Sigma}$ on 
$\mathbb{R}\times \Sigma$ and the hermitian structure on ${\rm pr}_{\Sigma}^{\ast}{\cal L}$; notice $D_A=(\frac{\partial}{\partial t}-iA_t)dt+d_{A(t)}$ here has a time component, and
we denote its curvature by $F_A=d(A_t dt +A)=:E_A\wedge dt + B_A$.
It was observed~\cite{Bog} that static configurations at critical 
self-coupling $\lambda^2=1$ 
are minima of the energy defined by ${\cal AH}_{1}$ if and only if they
satisfy the first-order equations (\ref{Bogo1})--(\ref{Bogo2}).
The (potential) energy of a gauged vortex with vortex number $N$ is then 
$\pi N$.
Since these equations are easier to study than the second-order Euler--Lagrange equations of (\ref{abelHiggs}), one might hope to understand the dynamics of the abelian Higgs model from a study of gauged vortices,
at least in the setting where
the velocities are small, so that the field configurations are
well approximated at any instant by solutions of the Bogomol'ny\u\i\ equations~\cite{ManRk}. One way to describe this so-called adiabatic approximation is as follows: 
construct from (\ref{abelHiggs}) an action on each $T{\cal M}_N$ 
by taking the fields to be solutions $(d_{A(z)},\Phi(z))$
of (\ref{Bogo1})--(\ref{Bogo2}) with time-dependent moduli 
$(z_{1}(t),\ldots,z_N(t))$, and integrate over $\Sigma$ in the $L^2$ norms.
It has been proven that the resulting mechanical system does give a good
description of 
the true (infinite-dimensional) vortex dynamics, even when we
shift slightly from critical self-coulpling~\cite{StuAH}. Upon this process of adiabatic reduction, a potential term in the field theory becomes
a potential function for the dynamics on ${\cal M}_N$.

The approximated  abelian Higgs dynamics on the moduli space was 
discussed by Samols~\cite{SamVS} following work by Strachan~\cite{StrLSV}.
It consists of geodesic motion with inertial mass $\pi$ (the static energy of one vortex) for 
a metric $g_{r\bar{s}}$ on ${\cal M}_N$. This metric is K\"ahler
with respect to the complex structure on ${\cal M}_N$ induced by the 
one on $\Sigma$. 
We call it the $L^2$ metric on ${\cal M}_N$, since it is obtained from the
natural $L^2$ norms (\ref{abelHiggs}) on the space of (covariant) derivatives 
of pairs $(D_A,\Phi)$. It is
described by the closed $(1,1)$-form
\begin{equation}\label{Samols}
\omega=\frac{i}{2}\sum_{r,s=1}^{N}g_{r\bar{s}}dz_{r}\wedge d\bar{z}_{s}=
\frac{i}{2}\sum_{r,s=1}^{N}
\left( {\Omega^2(z_{r}) \delta_{rs}}+
2\frac{\partial \bar{b}_s}{\partial z_r}
\right)dz_{r}\wedge d\bar{z}_s.
\end{equation}
Here, $z_r$ are the moduli associated to a complex coordinate $z$
for which $ds^{2}_{\Sigma}=\Omega^2(z) |dz|^2$. The quantities $b_r(z_{1},\ldots,z_N)$ are defined as follows. One can combine 
(\ref{Bogo1})--(\ref{Bogo2}) into a single equation for 
the gauge-invariant quantity $h:=\log \langle \Phi,\Phi\rangle$,
\begin{equation}\label{Taubes}
4\frac{\partial^{2}h}{\partial{z}\partial{\bar{z}}}+\Omega^{2}(z)(1-e^{h})
=4 \pi \sum_{r=1}^{N}\delta(z-z_r),
\end{equation}
where $\delta$ is the Dirac delta-function. 
A solution $h(z;z_1,\ldots,z_N)$ to (\ref{Taubes}) has an expansion
\begin{equation}\label{expansion}
h(z)=\log|z-z_r|^2+a_r
+\frac{1}{2}b_{r}(z-z_r)+
+\frac{1}{2}\bar{b}_{r}(\bar{z}-\bar{z}_r)+
\frac{\Omega^2(z_r)}{4}|z-z_r|^2+\cdots
\end{equation}
about a simple zero $z_r$ of $\Phi$. It is a remarkable fact that only the
linear coefficients $b_r$ in this expansion appear in~(\ref{Samols}).
It is clear that (\ref{expansion}) only makes sense in a neighbourhood
of the vortex $z_r$ which does not contain any other vortex positions.
In fact, (\ref{Taubes}) implies that we can write~\cite{SamVS} for each $r=1,\ldots,N$
\begin{equation}\label{bsing}
b_r(z_1,\ldots,z_N)=\sum_{s=1\atop{ s\ne r}}^{N}\frac{2}{z_r-z_s}+
\tilde{b}_r(z_1,\ldots,z_N),
\end{equation}
where $\tilde{b}_r$ are smooth on the coincidence locus $\Delta$.
In an expansion about the position of a vortex with multiplicity $n\le N$, the logarithmic term in (\ref{expansion}) has a prefactor $n$. The local 
moduli $z_r$ cannot be used to describe a neighbourhood in ${\cal M}_N$ of 
such a vortex configuration, and so the coefficients in an expansion 
equivalent to (\ref{expansion}) cannot be expressed as functions of the
$z_r$. 

Samols's formula (\ref{Samols}) still does not give the K\"ahler form explicitly, since the nontrivial quantities $b_r$ are specified in terms of 
unknown solutions to (\ref{Taubes}). Extracting concrete information about the moduli space metrics is still a nontrivial challenge. One may feel that work in this direction must inevitably rely on a numerical study of equation~(\ref{Taubes}); however, analytical results have been derived in particular cases using approximations of some sort~\cite{ManSpeAI, BatMan}, integrability~\cite{StrLSV}, or even a remarkable argument involving T-duality in string theory~\cite{TonTD}. 
In the following, we shall obtain further analytical information about the moduli space metrics when $\Sigma$ is a 2-sphere.

\section{Vortex dynamics in a background potential}

In this paper, we would like to extend the abelian Higgs model to include
a coupling with a background potential, which we define to be any smooth function $f:\Sigma \rightarrow \mathbb{R}$. 
To introduce the coupling term, we first define a vorticity 2-form 
$v(D_A,\Phi)$ on $\Sigma$ by the equation (cf.~\cite{ManNasCL})
\[
(d\,j+F_{A})\wedge dt =: v \wedge dt,
\]
where $j$ is the gauge-invariant supercurrent 1-form on 
$\mathbb{R}\times\Sigma$
\[
j:={\rm Im}\, \langle \phi, D_{A}\phi  \rangle.
\]
The coupling to the potential $f$ we propose is given by adding to the
potential energy term of the action
functional (\ref{abelHiggs}) the interaction term
\begin{equation}\label{coupling}
\mu^2 \int_{\Sigma}f\,v
\end{equation}
where $\mu^2$ is a coupling constant.

The Euler--Lagrange equations for the modified action 
\begin{equation} \label{model}
{\cal AH}_{\lambda^2}(D_A,\Phi)-\mu^2\int f\,v(D_A,\Phi)\wedge dt
\end{equation}
are as follows. We obtain the same
Gau\ss's law as for the abelian Higgs model (\ref{abelHiggs}),
\begin{equation} \label{Gauss}
d\ast E_A = 2 \ast {\rm Im}\,\langle \Phi, D_t \Phi \,dt\rangle
\end{equation}
(where $D_t:=\frac{\partial}{\partial t}-iA_t$),
which is to be regarded as a constraint on the space of fields determining $A_t$. The dynamical equations are
\begin{eqnarray}
&\D d\ast B_A-\ast \frac{\partial E_A}{\partial t} =-\ast \,{\rm Im}\,\langle \Phi, d_A \Phi\rangle -
\mu^2 (\langle \Phi, \Phi\rangle+1) \, df\label{Max},&\\
& \D \Box \Phi=\frac{\lambda^2}{2} (\langle\Phi, \Phi\rangle-1)\Phi+2i\mu^2 \ast (df\wedge d_A \Phi),& \label{nlKG}
\end{eqnarray}
where $\Box:=(D_{t})^{2} -\ast \, d_A \ast d_A$ is the covariant d'Alembertian on ${\rm pr_{\Sigma}^{*}}{\cal L}\rightarrow \mathbb{R}\times\Sigma$.
As expected, there are new terms (proportional to $\mu^2$) adding to the 
current in Amp\`ere's law (\ref{Max}), and also to the potential in the nonlinear Klein--Gordon equation (\ref{nlKG}). 

From our discussion in section \ref{secvortices}, we know that solutions to the vortex equations (\ref{Bogo1})--(\ref{Bogo2}) solve the equations of motion 
(\ref{Gauss})--(\ref{nlKG}) above in the static case (and setting $A_{t}=0$), provided $\lambda^2 =1$ and $\mu^2=0$. When the couplings are perturbed slightly away from these critical values, we expect a slow-moving solution to have a best vortex approximation, and that we can follow its evolution under the dynamics defined above. A detailed analysis of the case $\lambda^2 \simeq 1$, $\mu^2=0$ has been carried out by Stuart in \cite{StuAH}, where distances to best vortex approximations were estimated (in terms of $\lambda^2-1$ and initial errors) after finite-time evolution, and shown to be controlled for evolution times of order ${O}\left(|\lambda^2-1|^{-1/2}\right)$. In this paper, we make the natural assumption that an analogous result holds for $\lambda^2=1$ and $\mu^2 \simeq 0$. This assumption can be regarded as a motivation to the study of the dynamics on the moduli space of gauged vortices with a certain 
(and rather natural) background potential.

The coupling (\ref{coupling}) satisfies a number of desirable properties. 
One of them is that, if we take $f$ to be a constant, it reduces 
to a constant potential on the moduli space. To see this, we use 
(\ref{Bogo1}) and (\ref{Bogo2}) to rewrite~\cite{ManNasCL}
\begin{eqnarray*}
v&=& -i\langle d_{A} \Phi,d_A \Phi \rangle + (1-\langle \Phi , \Phi \rangle)B_A\\
&=& -i\langle \partial_{A}\Phi ,\partial_{A}\Phi\rangle+2(\ast B_A)B_A,
\end{eqnarray*}
which is precisely twice the energy density 
for solutions of the Bogomol'ny\u\i\ equations at critical self-coupling. 
Thus
\[
\int_{\Sigma} v|_{ {\cal M}_N}=2 \pi N.
\]
Notice that $df=0$ in this situation, so the terms proportional to $\mu^2$ in the field equations (\ref{Max}) and (\ref{nlKG}) vanish. This is a rather degenerate case of our model (\ref{model}), but the fact that the adiabatic approximation leads to sensible results at this level is already reassuring.

For the rest of the paper, we shall restrict ourselves to vortices living on a
2-sphere of radius $R$, $\Sigma=S^{2}_{R}$. In this context, we shall illustrate 
another natural property of the interaction term (\ref{coupling})
in section~\ref{secrotsym}.

\section{Localisation and the partition function} \label{seclocal}

The exact results we want to derive refer to the case $\Sigma = S^2_{R}$, on which $z$ shall denote a stereographic coordinate. The metric on $S^2_{R}$
has K\"ahler (volume) form
\begin{equation}\label{omegaS2R}
\omega_{S^{2}_{R}} = \frac{2 i R^2}{(1+|z|^2)^2} dz \wedge d\bar{z}
\end{equation}
and the constraint (\ref{Bradlow}) reads 
\[
R^2>N.
\]

The moduli spaces in this case are
\begin{equation}\label{MNCPN}
{\cal M}_{N}={\rm Sym}^{N}(S^{2}_R) \cong \mathbb{CP}^N,
\end{equation}
equipped with the K\"ahler structures
\begin{equation}\label{omegaMS2R}
\omega={i}\sum_{r,s=1}^{N}
\left( \frac{R^2 \delta_{rs}}{(1+|z_r|^2)^2}+
\frac{\partial \bar{b}_s}{\partial z_r}
\right)dz_{r}\wedge d\bar{z}_s.
\end{equation}
One way to understand (\ref{MNCPN}) is as follows. Let 
$V_{j}\subset {\cal M}_N$ be the locus where precisely $j$ vortices are
at $z=\infty$. Then $V_j$ is parametrised by the (unordered) positions of
the remaining $N-j$ vortices on $\mathbb{C}$, which are 
unambiguously specified by the coefficients of a monic polynomial of degree $N-j$ having these positions
as roots. Hence $V_{j}\cong\mathbb{C}^{N-j}$. The way in which these 
$2(N-j)$-cells are glued together in ${\cal M}_N$ is determined by 
attaching maps corresponding to letting one vortex go to $z=\infty$ at
a time. This yields precisely the description of $\mathbb{CP}^N$ as a 
CW-complex.

\subsection{Rotational symmetry}\label{secrotsym}

For most of our discussion, we shall restrict our attention to the potential
\begin{equation} \label{Rx3}
f 
= R^2 \frac{1-|z|^2}{1+|z|^2}.
\end{equation}
This potential is very natural if we assume that there is a special axis on the
sphere. In fact, (\ref{Rx3}) is the simplest nontrivial circularly symmetric function on $S^{2}_{R}$, in the sense that other potentials with circular symmetry can be expanded as power series in $f$.

It is easy to check that (\ref{Rx3}) is a hamiltonian for the circle action 
of rotations around the axis of $S^{2}_R$ associated to the stereographic
coordinate $z$,
\[
\iota_{i\left(z\frac{\partial}{\partial z}-\bar{z}\frac{\partial}{\partial \bar{z}}\right)}\omega_{S^2_R} = -df.
\]
There is an induced circle action on 
${\cal M}_{N}={\rm Sym}^{N}(S^{2}_R)$ with generator
\begin{equation}\label{xi}
\xi=i\sum_{r=1}^{N}\left( z_r \frac{\partial}{\partial z_r} -
\bar{z}_{r}\frac{\partial}{\partial \bar{z}_r}\right)
\end{equation}
which extends to a smooth vector field on ${\cal M}_{N}$. 
Rotational symmetry on $S^{2}_{R}$ implies \cite{RomQCS}
\begin{equation} \label{rotsymid}
\sum_{r=1}^{N}\left(z_r b_r - \bar{z}_r \bar{b}_r\right)=0,
\end{equation}
and one can use this equality to show that the K\"ahler structure (\ref{omegaMS2R}) is preserved by the one-parameter group generated by (\ref{xi}). Since $\omega$ is closed, it follows from
\[
0={\rm \pounds}_{\xi}\omega=\iota_{\xi}(d\omega)+d(\iota_{\xi}{\omega})=
d(\iota_{\xi}{\omega})
\]
and $H^{1}(\mathbb{CP}^N)=0$ that there is also a hamiltonian for the circle
action on ${\cal M}_N$, which can be computed to be
\begin{equation}\label{J}
J=2 \pi \sum_{r=1}^{N}\left( R^2 \frac{1-|z_r|^2}{1+|z_r|^2}
-(z_r b_r +1) \right).
\end{equation}
Notice that this formula assumes that all the vortices are separated,
but it does extend to a smooth function on the whole of ${\cal M}_N$.

Hamiltonians of circle actions are defined up to a constant. However, this
constant is fixed if the circle involved is a one-parameter subgroup of a
Lie group with discrete centre and hamiltonian action on the symplectic manifold, and we demand
that the hamiltonian should be a component of the corresponding moment map.
In our case, the circle action extends to a hamiltonian
action of ${\rm Iso}(S^2_R)={\rm SO}(3)$ on ${\cal M}_{N}$. The choice
of constants in (\ref{J}) can be checked~\cite{RomQCS} to be consistent with the moment map ${\cal M}_{N}\rightarrow \mathfrak{so}(3)^{\ast}$. 
The same can be said of $f$ in (\ref{Rx3}) in relation to the symplectic structure $\omega_{S^2_R}$ in (\ref{omegaS2R}).

A natural question to ask at this point is whether our coupling behaves well
with respect to moment maps. We can regard the adiabatic reduction of couplings of the form (\ref{coupling}) as defining a linear map between Lie algebras
\begin{equation} \label{defR}
\begin{array}{rrcl}
{\cal R}: &  C^{\infty}(S^{2}_{R}, \omega_{S^2_R}) 
& \longrightarrow & C^{\infty}({\cal M}_N,\omega) \\
 & f & \longmapsto & \int_{S^2_R}f\,v
\end{array}
\end{equation}
for each $N$ (where the Lie bracket on each side is the Poisson bracket defined by the symplectic structure). One may hope that this map preserves the
Lie algebra structures, and that it relates corresponding components of the ${\rm SO}(3)$-moment maps on each of $S^2_R$ and ${\cal M}_N$. The 
following proposition shows that this is indeed the case.

\begin{proposition}\label{couplandmu}
For each $N<R^2$, there is a commutative diagram of Lie algebras
\[
\begin{array}{rrc}
&&C^{\infty}(S^2_R)\\
& {\nearrow} & \\
\mathfrak{so}(3) & & \Big{\downarrow} \; {\cal R}\\
& {\searrow} & \\
&&C^{\infty}({\cal M}_N)\\
\end{array}
\]
where the diagonal arrows denote the dual maps to the moment 
maps $S^2_R \rightarrow \mathfrak{so}(3)^{\ast}$ and 
${\cal M}_N \rightarrow \mathfrak{so}(3)^{\ast}$, respectively.
\end{proposition}
\begin{proof}
The existence of the moment maps is guaranteed by the vanishing of the 
Lie algebra cohomology group $H^{2}(\mathfrak{so}(3);\mathbb{R})$, 
a consequence of simplicity~\cite{GuiSte}. 

Given the linearity of ${\cal R}$ in (\ref{defR}), the proposition will follow if we 
check the commutativity of the diagram on the generators of 
$\mathfrak{so}(3)$.

We start by sketching how to obtain 
\begin{equation}\label{reduction}
J={\cal R}(f), 
\end{equation}
where $f$ is given by (\ref{Rx3}).
This calculation is paradigmatic of the process of 
reduction to the moduli space. 
Suppose that the vortices are all separated, and work first on the subset
\[
C_{\epsilon}:=
\left\{z\in \mathbb{C}: |z|<\frac{1}{\epsilon} \right\}-\bigcup_{r=1}^{N}B_{\epsilon}(z_r)\;
\subset \;S^{2}_R,
\]
where $\epsilon$ is taken small enough.
Then (\ref{Taubes}) can be used to write on $C_\epsilon$
\[
v=\frac{i}{2R^2}\,d\left( (1+|z|^2)^2 \frac{\partial^2 h}
{\partial z \partial \bar{z}}(\bar{\partial} -\partial)h\right)
\]
and
\[
fv=\frac{i}{2R^2}\,d\left(
\frac{\partial}{\partial z}(1-|z|^4)\frac{\partial}{\partial z}
\left( \frac{\partial h}{\partial \bar{z}}\right)^2+
\frac{\partial}{\partial z}(1+|z|^2)^2
\left( \frac{\partial h}{\partial \bar{z}}\right)^2
\right)\wedge d\bar{z}.
\]
Using Stokes' theorem and the estimates
\[
\frac{\partial^{2}h}{\partial z \partial\bar{z}}=
-\frac{R^2}{(1+|z_r|^2)^2}+o(\epsilon),\quad
\frac{\partial h}{\partial\bar{z}}=\frac{1}{\epsilon}e^{i \arg(z-z_r)}
+\frac{\bar{b}_r}{2}+o(1)\qquad {\rm as}\;\epsilon \rightarrow 0 
\]
for $z\in \partial B_{\epsilon}(z_r)$,
which follow from (\ref{expansion}), we do obtain (\ref{reduction}) 
after taking $\epsilon\rightarrow 0$.

To proceed, we can either repeat the calculation for the other generators
(as in~\cite{RomQCS}), keeping track of the constants to preserve the
$\mathfrak{so}(3)$ algebra, or change integration variables and apply 
the appropriate rotations to the
two sides of (\ref{reduction}). The second procedure becomes straightforward
once we observe that the $b_r$ transform as~\cite{ManNasVVM}
\[
(T^{\ast}b_{r})(z_1,\ldots z_N)=
\frac{1}{T'(z_r)}b_r(z_1,\ldots,z_N)-\frac{T''(z_r)}{T'(z_r)^2} 
\]
under any holomorphic $T \in {\rm Iso}(\Sigma)$. This equation is readily 
obtained from the expansion (\ref{expansion}).

\end{proof}

It should be noted that ${\cal R}$ does not preserve the structures of 
$C^\infty(S^{2}_{R},\omega_{S^2_{R}})$ and 
$C^\infty({\cal M}_N,\omega)$ as Poisson algebras.

Since the moduli spaces are compact in our case, the circle actions we are interested in must have fixed points (the zeroes of $\xi$). It is easy to see that these are exactly the $N+1$ points $p_{j}\in {\cal M}_{N}$ (with $j=0,1,\ldots, N$) describing configurations of $j$ vortices at $z=0$ and $N-j$ vortices at $z=\infty$, which will be fixed by a rotation of all the vortices around the axis through $0$ and $\infty$.
We remark that $J$ is a Morse function on ${\cal M}_{N}$, with 
critical set
\[
{\rm Crit}(J)=\{p_{0},p_{1},\ldots,p_{N}\}.
\]

In the next section, it will be useful to understand the circle action in
the neighbourhood of the fixed points. On each tangent space 
$T_{p_j}{\cal M}_{N}\cong\mathbb{C}^{N}$, the action linearises to a complex $N$-dimensional representation of the circle group; this in turn decomposes 
into $N$ $1$-dimensional representations of ${\rm U}(1)$, and each of them is uniquely specified by its weight $k\in \mathbb{Z}$:
\[
{e^{2\pi i t}}: \zeta \mapsto e^{2 \pi i k t}\zeta, \qquad t\in \mathbb{R}, \quad\zeta\in\mathbb{C}.
\]
Thus to each of the fixed points $p_{j}$ we can associate $N$ weights 
$k_{j,\ell}$, $1 \le \ell \le N$, well defined up to order. The product of
the weights at each fixed point
\begin{equation}\label{edefined}
e(p_{j}):=\prod_{\ell=1}^{N}k_{j,\ell}
\end{equation}
is a local invariant of the circle action: it does not depend on the 
choice of
coordinates on the moduli space. We shall make use of the following result:
\begin{lemma}\label{invepj} 
For the circle action generated by (\ref{J}) on $({\cal M}_N,\omega)$,
\[
e(p_{j})=(-1)^{j+N} j! (N-j)!,\qquad 0\le j \le N.
\]
\end{lemma}
\begin{proof}
In terms of the moduli $z_r$, the circle action is simply given by
\begin{equation} \label{caction}
e^{2 \pi i t}: z_{r}\mapsto e^{2 \pi i t} z_{r},\qquad t\in \mathbb{R}.
\end{equation}
However, all the $p_{j}$ except $p_{0}$ lie in $\Delta$, where the coordinate
system defined by the $z_{r}$ becomes singular, and so we must introduce other
coordinates to compute the weights. We fix $j$ and arrange the vortex labels
such that vortices $1,\ldots,j$ are at $z=0$, and vortices $j+1,\ldots,N$
are at $z=\infty$. (We are allowed to do this since the vortices in each
cluster are to be thought of as interchangeable, but vortices belonging to different clusters have separate identities.) Now we introduce
\[
\begin{array}{ll}
u_{j,r}:=s_{r}^{[j]}(z_1,\ldots,z_j), & \quad 1\le r \le j, \\
v_{j,r}:=s_{r}^{[N-j]}({z^{-1}_{j+1}},\ldots,{z^{-1}_{N}}), & \quad 1 
\le r \le N-j, 
\end{array}
\]
where $s_r^{[j]}$ denotes the $r$th elementary symmetric polynomial in
$j$ variables,
\[
s_{r}^{[j]}(t_{1},\ldots,t_j):=\sum_{i_{1}<\cdots<i_r}t_{i_1}\cdots t_{i_r}.
\]
Clearly, $\{u_{j,1},\ldots,u_{j,j},v_{j,1},\ldots,v_{j,N-j}\}$
is a centred local coordinate system at $p_j\in{\cal M}_N$, and we can also
use it as a coordinate system on $T_{p_j}{\cal M}_N$.
From (\ref{caction}), we find that the linearisation of the circle action
is described by
\begin{eqnarray*}
&&u_{j,r}\mapsto e^{2 \pi i r t}u_{j,r}\\
&&v_{j,r}\mapsto e^{2 \pi i (-r) t}v_{j,r}
\end{eqnarray*}
and therefore we obtain in (\ref{invepj})
\[
e(p_{j})=\left(\prod_{k=1}^{j}k\right)\left(\prod_{\ell=1}^{N-j}(-\ell)\right)
=(-1)^{j+N}j!(N-j)!.
\]
\end{proof}

\subsection{The partition function}

The dynamics on the moduli space of $N$ vortices is defined by the 
lagrangian
\[
L=\frac{\pi}{2} \sum_{r,s=1}^{N}g_{r\bar{s}}(z_1,\ldots,z_N) \dot{z}_{r}\dot{\bar{z}}_{s}-
\mu^2 J(z_1,\ldots,z_N),
\]
where $g_{r\bar{s}}$ are the coefficients of the metric on ${\cal M}_N$
and $\pi$ is the mass of a single vortex.
In the canonical picture, we describe the dynamics as a hamiltonian system 
on the $4N$-dimensional manifold $T^{\ast}{\cal M}_N$ equipped with its
canonical symplectic structure
\[
\omega_{\rm can}=\frac{1}{2}\sum_{r=1}^{N}
(dz_r\wedge d\bar{w}_r + d\bar{z}_r\wedge dw_r).
\]
Here, the $w_{r}$ are complex coordinates for the fibres of $T^{\ast}{\cal M}_{N}\rightarrow{\cal M}_{N}$, conjugate to the moduli $z_r$:
\[
w_{r}=\frac{\partial L}{\partial \dot{z_{r}}}=\pi \sum_{s=1}^{N}g_{r\bar{s}}\dot{\bar{z}}_s.
\]
The dynamics is generated by the hamiltonian
\[
H=\frac{1}{2\pi}\sum_{r,s=1}^{N}g^{r\bar{s}}(z_1,\ldots,z_N)w_r \bar{w}_s+
\mu^2 J(z_1,\ldots,z_N),
\]
where $g^{r\bar{s}}$ denote the entries of the inverse to the matrix of 
coefficients of the metric.

According to the canonical formalism for classical statistical mechanics, the partition function of the vortex moduli space dynamics is given by the Gibbs formula
\begin{equation}\label{Zbrute}
Z=\frac{1}{(2 \pi \hbar)^{2N}}\int_{T^{\ast}{\cal M}_N}
e^{-H(z_{1},\ldots,z_{N},w_{1},\ldots,w_N)/T}
\;\frac{\omega_{\rm can}^{2N}}{(2N)!}.
\end{equation}
The prefactor to the integral is Planck's constant $2\pi \hbar$ raised to 
the power $\frac{1}{2}\dim_{\mathbb{R}} T^{\ast}{\cal M}_{N}=2N$, and $T$ is the temperature. 
We are normalising Boltzmann's constant to unity.

In parallel with the calculation in \cite{ManSMV}, we find that the integral
(\ref{Zbrute}) factorises as a product of a gaussian integral along the fibres,
which can be readily calculated, and an integral over the moduli space with
the Liouville measure associated to (\ref{omegaMS2R}):
\begin{equation}\label{Zfactors}
Z=\left( \frac{T}{2\hbar^2}\right)^{N}\int_{{\cal M}_N}e^{-{\mu^2}
J(z_1,\ldots,z_N)/T}\;
\frac{\omega^{N}}{N!}.
\end{equation}
The integral remaining still looks very complicated, but we shall show 
that it can also be computed exactly, using localisation in symplectic 
geometry. The main tool we will use is the following version of a famous result by Duistermaat and Heckman \cite{DuiHec}:

\begin{theorem}[Duistermaat--Heckman formula]
Let $(M,\omega)$ be a $(2n)$-dimensional compact symplectic manifold
with a hamiltonian circle action generated by a Morse function 
$K:M\rightarrow \mathbb{R}$. Then
\begin{equation}\label{DuistHeck}
\int_{M}e^{\tau K}\;\frac{\omega^{n}}{n!}=\sum_{p\in{\rm Crit}(K)}
\frac{e^{\tau K(p)}}{\tau^{n}e(p)},
\end{equation}
where $e(p)$ denotes the product of the weights of the linearised action at
the critical point $p$, and $\tau$ is a formal parameter.  
\end{theorem}

A streamlined proof of this theorem can be found in section 5.6 of reference \cite{McDSal}.

In our context, taking $M={\cal M}_N$, $K=J$ and $\tau=-{\mu^2}/{T}$, and
making use of Lemma~\ref{invepj}, the equality (\ref{DuistHeck}) takes the 
form \begin{equation}\label{ourDH}
\int_{{\cal M}_N}e^{-\mu^2 J/T}\;\frac{\omega^N}{N!}=
\sum_{j=0}^{N}(-1)^{j}\frac{T^{N} e^{-\mu^2 J(p_{j})/T}}{\mu^{2N}j!(N-j)!}.
\end{equation}
Hence, to compute the integral in (\ref{Zfactors}), we only need to
evaluate the potential $J$ at the critical points $p_j \in {\cal M}_N$.

We shall make use of the spherical symmetry of the problem to determine 
the contribution
of the $\tilde{b}_r$ terms in the 
formula (\ref{J}) to $J(p_j)$. Suppose that the
Higgs field $\Phi$ has a zero at $z=y$ of order $j$ and a zero at 
$z=-\frac{1}{\bar{y}}$ of order $N-j$, a vortex configuration corresponding 
to a point of ${\cal M}_N$ that we shall denote by $p_{j}(y)$.
Then $h=\log\langle\Phi,\Phi\rangle$ must satisfy
\begin{equation}\label{Taubesanti}
\frac{\partial^2 h}{\partial{z}\partial{\bar{z}}}-
\frac{R^2 (e^h-1)}{(1+|z|^2)^2}=j \pi \delta(z-y)+(N-j)\pi\delta\left( z-\frac{1}{\bar{y}}\right).
\end{equation}
This equation leads to the following expansions for $h$: 
\[
h(z,y)=j\log|z-y|^2+a_{+}(y)+\frac{1}{2}b_{+}(y)(z-y)+
\frac{1}{2}\overline{b_{+}(y)}(\bar{z}-\bar{y})+\cdots
\]
around $z=y$, and
\[
h(z,y)=(N-j)\log\left|z+\frac{1}{\bar{y}}\right|^2+a_{-}(y)+
\frac{1}{2}b_{-}(y)\left(z+\frac{1}{\bar{y}}\right)+
\frac{1}{2}\overline{b_{-}(y)}\left(\bar{z}+\frac{1}{y}\right)+\cdots
\]
around $z=-\frac{1}{\bar{y}}$. 
We need to calculate $b_{\pm}(y)$.
Extending an argument in~\cite{ManSMV}, we explore the fact that
$h(z,y)$ must be a function of the spherical (and thus also of the chordal) distance between the points of coordinates $z$ and $y$ on $S^{2}_{R}$. The square of the chordal distance is given by
\[
\frac{4R^2|z-y|^2}{(1+|z|^2)(1+|y|^2)},
\]
and so $h$ can be expanded as
\begin{equation}\label{hzeqy}
h(z,y)=j\log\frac{|z-y|^2}{(1+|z|^2)(1+|y|^2)}+c_{+}+
d_{+}\frac{|z-y|^2}{(1+|z|^2)(1+|y|^2)}+\cdots.
\end{equation}
For $y=0$, this yields
\[
h(z,0)=j\log|z|^2+c_{+}+(d_{+}-j)|z|^2+\cdots;
\]
substituting this into (\ref{Taubesanti}) with $y=0$ and small $z$, and 
looking at the zeroth order term in $|z|^2$, we conclude that
\[
d_{+}=j-R^2.
\]
Now taking $y$ arbitrary, we rearrange (\ref{hzeqy}) as
\begin{eqnarray*}
h(z,y)&\!\!=\!\!&j\log|z-y|^2+c_{+}
-2j\log(1+|y|^2)-\frac{j\bar{y}}{1+|y|^2}(z-y)-
\frac{jy}{1+|y|^2}(\bar{z}-\bar{y})\\
&&+\frac{j\bar{y}^2}{(1+|y|^2)^2}(z-y)^2+
\frac{jy^2}{(1+|y|^2)^2}(\bar{z}-\bar{y})^2-
\frac{R^2}{(1+|y|^2)^2}|z-y|^2+\cdots
\end{eqnarray*}
to read off
\begin{equation}\label{bofy}
b_{+}(y)=-\frac{2j\bar{y}}{1+|y|^2}.
\end{equation}
We proceed similarly for $b_{-}(y)$: writing
\begin{equation}\label{hzeqanty}
h(z,y)=(N-j)\log\frac{\left| z+\frac{1}{\bar{y}}\right|^2}
{(1+|z|^2)\left( 1+\frac{1}{|y|^2}\right)^2}+c_{-}+
d_{-}\frac{\left| z+\frac{1}{\bar{y}}\right|^2}
{(1+|z|^2)\left( 1+\frac{1}{|y|^2}\right)^2}+\cdots
\end{equation}
we can calculate
\[
c_{-}=N-j-R^2
\]
by taking $y=\infty$, substituting in (\ref{Taubesanti}) and looking
at the zeroth term in the large $|z|^2$ expansion; then rearrange
(\ref{hzeqanty}) as
\begin{eqnarray*}
h(z,y)&\!\!=\!\!&
(N-j)\log\left|z+\frac{1}{\bar{y}}\right|^2-2(N-j)\log\frac{1+|y|^2}{|y|}
+c_{-}\\
&& +\frac{(N-j)\bar{y}}{1+|y|^2}\left(z+\frac{1}{\bar{y}}\right)
+\frac{(N-j){y}}{1+|y|^2}\left(\bar{z}+\frac{1}{y}\right)\\
&&+ \frac{(N-j)\bar{y}^2}{(1+|y|^2)^2}\left(z+\frac{1}{\bar{y}}\right)^2
+\frac{(N-j){y}^2}{(1+|y|^2)^2}\left(\bar{z}+\frac{1}{y}\right)^2
-\frac{R^2|y|^2}{(1+|y|^2)^2}\left|z+\frac{1}{\bar{y}}\right|^2+\cdots
\end{eqnarray*}
and read off
\begin{equation}\label{Bofy}
b_{-}(y)=\frac{2(N-j)\bar{y}}{1+|y|^2}.
\end{equation}

To proceed, we must interpret the formula (\ref{J}) carefully to deal with vortex clusters
such as the configurations $p_{j}(y)\in{\cal M}_N$. When any number of vortices
become coincident, the smooth part $\tilde{b}_r$ of their individual $b_{r}$ coefficients tends to the linear coefficient in the cluster expansion of $h$ about the position of the cluster, but the singularities in (\ref{bsing}) must be treated with some care.
Although these singular parts diverge separately, they always yield a finite contribution in a $\mathfrak{S}_N$-invariant linear combination over the individual vortices --- for example, they cancel mutually in a term like $\sum_{r=1}^{N}b_{r}$. In (\ref{J}), these singular parts are multiplied by the vortex positions, and so in the 
clustering $(z_1,\ldots,z_{N})\rightarrow p_j(y)$ they give a contribution
\begin{eqnarray}
-2\pi \sum_{r=1}^{N}z_{r}
\sum_{s\neq r \atop {\rm same\; cluster}}
\frac{2}{z_r - z_s}&=&
-4\pi\sum_{r=1}^{j}
\sum_{s\neq r}
\frac{z_r}{z_r - z_s}
-4\pi\sum_{r=j+1}^{N}
\sum_{s\neq r }
\frac{z_r}{z_r - z_s} \nonumber \\
&=&
-4\pi\left( \sum_{r=1}^{j} \sum_{s< r}+\sum_{r=j+1}^{N}\sum_{s< r}\right)
\left(\frac{z_r}{z_r - z_s}+\frac{z_s}{z_s- z_r} \right) \nonumber\\
&=&
-4\pi\left( \sum_{r=1}^{j}\sum_{s<r} +\sum_{r=j+1}^{N}\sum_{s<r}\right)1 \nonumber \\
&=&
-4\pi\left( 
\frac{j(j-1)}{2}+\frac{(N-j)(N-j-1)}{2} \right) \nonumber \\
&=&
-2\pi((N-j)^2 +j^2 -N) 
\label{contrsing}
\end{eqnarray}
to $J(p_{j}(y))$. The remaining part of $J(p_{j}(y))$ is
\begin{eqnarray} 
2 \pi\sum_{r=1}^{N}
\left(R^2 \frac{1-|z_r|^2}{1+|z_r|^2}-(z_r \tilde{b}_r+1) \right)
&=&
2\pi R^2\left(
j\frac{1-|y|^2}{1+|y|^2}+(N-j)\frac{1-\frac{1}{|y|^2}}{1+\frac{1}{|y|^2}}
\right) \nonumber\\
&&
-2\pi\left(
j(y\, b_{+}(y)+1)+(N-j)\left(\left(-\frac{1}{\bar{y}}\right)b_{-}(y)+1\right)
\right) \nonumber \\
&=& 2\pi(2j-N)(R^2 -N)\frac{1-|y|^2}{1+|y|^2} \nonumber\\
&&
+2\pi((N-j)^2+j^2-N), \label{contrnonsing}
\end{eqnarray}
where we used (\ref{bofy}) and (\ref{Bofy}).
Adding (\ref{contrsing}) and (\ref{contrnonsing}), we obtain 
\[
J(p_{j}(y))=2 \pi (R^2-N)(2j-N)\frac{1-|y|^2}{1+|y|^2}
\]
and hence
\begin{equation}\label{Jpj}
J(p_j)=J(p_j(0))=2\pi(R^2 - N)(2j-N),\qquad 0\le j\le N.
\end{equation}

Inserting (\ref{Jpj}) in (\ref{ourDH}), we find 
\begin{equation}\label{intMN}
\int_{{\cal M}_N}e^{-\mu^2 J/T}\;\frac{\omega^N}{N!}=
\sum_{j=0}^{N}(-1)^j\frac{T^N 
e^{-2 \pi \mu^2 (R^2 -N)(2j-N)/T}}{\mu^{2N}j!(N-j)!}.
\end{equation}
This equation can be interpreted as an equality in the formal power series 
ring $\mathbb{R}[[\frac{\mu^2}{T}]]$, which is equivalent to an infinite 
number of identities over the reals. The first $N$ nontrivial identities
are
\begin{equation}\label{ident0}
\sum_{j=0}^{N}\frac{(-1)^j}{j!(N-j)!}(2j-N)^{k}=0,\qquad 0\le k\le N-1,
\end{equation}
and they must be true for consistency. But they are implied by the following technical lemma.

\begin{lemma}\label{combinat}
For all $N \in \mathbb{N},$
\[
\sum_{j=0}^{N}\frac{(-1)^j}{N!}{N \choose j}
\left(\frac{N}{2}-j \right)^{k} =
\left\{
\begin{array}{r@{\quad{\rm if}\quad}l}
0 & 0 \le k \le N-1, \\
1 & k=N.
\end{array}
\right.
\]
\end{lemma}

\begin{proof}
We start from the equality
\[
\left(x \frac{d}{dx}\right)^{\ell}(1-x)^{N}=\sum_{j=0}^{N}(-1)^j
{N \choose j}j^{\ell}x^{j},\qquad \ell \in \mathbb{N}\cup\{0\},
\]
obtained by successively acting on the binomial expansion with the 
Euler operator $x\frac{d}{dx}$.  Setting $x=1$ yields
\[
\sum_{j=0}^{N}(-1)^j{N \choose j} j^{\ell}=
\left.\left( x\frac{d}{dx}\right)^{\ell}(1-x)^N\right|_{x=1}.
\]
We use this to write
\begin{eqnarray}
\sum_{j=0}^{N}(-1)^j {N \choose j}\left( \frac{N}{2}-j\right)^k &=&
\sum_{j=0}^{N}\sum_{\ell=0}^{k}(-1)^{j+\ell}{N \choose j}{k \choose \ell}
\left(\frac{N}{2}\right)^{k-\ell}j^{\ell}\nonumber \\
&=& \sum_{\ell=0}^{N}(-1)^{\ell}{k \choose \ell}\left(\frac{N}{2}
\right)^{k-\ell}\sum_{j=0}^{N}(-1)^{j}{N \choose j}j^{\ell}\nonumber\\
&=& \sum_{\ell=0}^{N}(-1)^{\ell}{k \choose \ell}\left(\frac{N}{2}
\right)^{k-\ell}\left.\left( x\frac{d}{dx}\right)^{\ell}(1-x)^N
\right|_{x=1}\nonumber \\
&=&\left.\left(\frac{N}{2}- x\frac{d}{dx}\right)^{k}(1-x)^N\right|_{x=1}.
\label{clearmess}
\end{eqnarray}
It is clear that (\ref{clearmess}) is zero whenever $0\le k<N $, since the
differential operator $(\frac{N}{2}-x\frac{d}{dx})^{k}$ will not annihilate
enough $(1-x)$ factors before $x\rightarrow 1$. However, if $k=N$ one obtains
\begin{eqnarray*}
\left.\left(\frac{N}{2} -x\frac{d}{dx}\right)^{N}(1-x)^{N}\right|_{x=1}&=&
\left.(-1)^{N}\left(x \frac{d}{dx}\right)^{N}(1-x)^N\right|_{x=1}\\
&=& \left.(-1)^{N}x^{N}\left(\frac{d}{dx}\right)^{N}(1-x)^{N}\right|_{x=1}\\
&=& N!
\end{eqnarray*}
making use of $[\frac{d}{dx},x]=1$.
\end{proof}

Notice that Lemma~\ref{combinat} yields yet another identity from (\ref{intMN}):
\begin{eqnarray}
{\rm Vol}({\cal M}_N) := \int_{{\cal M}_N}\frac{\omega^N}{N!}
& = &\frac{(4 \pi)^N (R^2 -N)^N}{N!}
\sum_{j=0}^{N}\frac{(-1)^j}{j!(N-j)!}\left( \frac{N}{2}-j\right)^N
\nonumber\\
& = & \frac{(4 \pi)^N(R^2 -N)^N}{N!}.
\label{VolMN}
\end{eqnarray}
This is precisely the formula found by Manton for the volume of the
vortex moduli space in \cite{ManSMV}, using a more direct
argument involving the cohomology of $\mathbb{CP}^N$. Equation (\ref{VolMN}) provides a nontrivial check of our calculations.

Beyond the identities~(\ref{ident0}) and the formula (\ref{VolMN}) for 
${\rm Vol}({\cal M}_N)$, our localisation argument yields an
infinite number of integrals over the moduli space: for
$m\in\mathbb{N}$
\begin{equation}\label{nontrivialints}
\int_{{\cal M}_N} J(z_1,\ldots,z_N)^m\;\frac{\omega^N}{N!}=
\sum_{j=0}^{N}\frac{(-1)^{N-j}m!}{j!(N-j)!(N+m)!}
\left( 2\pi(R^2 -N) (2j-N)\right)^{N+m}.
\end{equation}
Notice that both sides of this equation vanish when $m$ is odd:
The left-hand side from reflection symmetry on $S^{2}_{R}$ and 
Proposition~\ref{couplandmu}, and the right-hand side from
\[
\sum_{j=0}^{N}\frac{(-1)^{j}}{j!(N-j)!}\left(\frac{N}{2}-j\right)^{N+2n-1}=0,
\qquad \forall n\in\mathbb{N},
\]
which follows from substituting $j$ by $N-j$ in the sum.
For $m$ even, (\ref{nontrivialints}) yields new quantitative information
about the metric on ${\cal M}_N$. 
We note in passing that even the vanishing integrals ($m$ odd) have
interesting content.
For example, using the result for $m=1$, we find from (\ref{J}) that
\[
\int_{{\cal M}_N}\sum_{r=1}^{N}z_r b_r(z_1,\ldots, z_N) \, \frac{\omega^N}{N!}=-{\rm Vol}({\cal M}_N);
\]
a similar argument for rotations around $z=1$ and $z=i$ yields
\[
\int_{{\cal M}_N}\sum_{r=1}^{N}b_r(z_1,\ldots,z_N)\, \frac{\omega^N}{N!}=0,
\]
which in turn also leads to
\[
\int_{{\cal M}_N}\sum_{r=1}^{N}z_{r}^2 b_r(z_1,\ldots,z_N)\, \frac{\omega^N}{N!}=0
\]
by a property of the functions $b_r$ 
analogous to (\ref{rotsymid}) \cite{RomQCS}:
\[
\sum_{r=1}^{N}(2 z_r + z^{2}_{r} b_r + \bar{b}_r)=0.
\]

Using the result (\ref{nontrivialints}), together with (\ref{VolMN}),
we can obtain the integral over ${\cal M}_N$ of any power series
in $J$. Such power series, when convergent, give analytic functions on
${\cal M}_N$ which are invariant under the circle action generated by~(\ref{xi}).

Finally, we can use (\ref{intMN}) to compute the partition function 
(\ref{Zfactors}) of the vortex gas 
in the backgound field (\ref{Rx3}) to be
\begin{eqnarray}
Z&=& 
\left( \frac{T^2}{2 \hbar^2 \mu^2}\right)^N 
e^{2\pi\mu^2 N(R^2 -N)/T}
\sum_{j=0}^{N}\frac{(-1)^j}{j!(N-j)!}e^{-4 \pi \mu^2(R^2-N)j/T}\nonumber \\
&=&
\frac{1}{N!}
\left( \frac{T^2}{2 \hbar^2 \mu^2}\right)^N 
e^{2\pi\mu^2 N(R^2 -N)/T}
\left(1-e^{-4\pi \mu^2 (R^2-N)/T}\right)^N \nonumber \\
&=&
\frac{1}{N!}
\left( \frac{T}{\hbar \mu}\right)^{2N} 
\sinh^{N} \left( \frac{2\pi \mu^2 (R^2-N)}{T} \right) .\label{Zresult}
\end{eqnarray}

\section{Thermodynamics of the vortex gas} \label{secvdW}

The Helmholtz free energy of the vortex system 
can be computed from (\ref{Zresult}) as
\[
F=-T\log Z \simeq -NT\left( 
\log \sinh \frac{\mu^2 (A-4\pi N)}{2T}-\log N + 2\log \frac{\sqrt{e}T}{\hbar \mu}
\right),
\]
where we made use of Stirling's approximation $\ln N! \simeq N \ln N - N$,
and we introduced the area of $S_{R}^{2}$, $A=4\pi R^2$. The entropy is
given by
\[
S=-\frac{\partial F}{\partial T}=N\left(\log \sinh \frac{\mu^2 (A-4 \pi N)}{2T}
+\log \frac{e^3 T^2}{\hbar^2 \mu^2 N}-\frac{\mu^2 (A-4\pi N)}{2T}
{\coth} \frac{\mu^2 (A-4\pi N)}{2T}\right).
\]
Both these quantities turn out to be nonextensive, due to a nonlinear effect
produced by the interaction with the external potential.

The interaction
is controlled by the coupling $\mu^2$; at small coupling, keeping $A$ and $N$
finite, we can approximate the hyperbolic functions to first order as $\sinh \chi \simeq \chi $ and $\coth \chi \simeq \frac{1}{\chi}$, leading to the same results found by Manton in 
the absence of interaction~\cite{ManSMV}. In this noninteracting setting, both $F$ and $S$ become extensive in the thermodynamical limit of $A\rightarrow \infty$, $N\rightarrow \infty$ and constant density $n=\frac{N}{A}$.
The pressure $P=-\frac{\partial F}{\partial A}$ of the system in this regime 
can be readily computed and yields the equation of state~\cite{ManSMV} 
\begin{equation}\label{Clausius}
P(A-4\pi N)=NT.
\end{equation}
This is a particular limit of the van der Waals equation, known as a Clausius
equation of state.
It holds more generally on any compact Riemann 
surface~\cite{ManNasVVM}. The fact that the factor $A-4\pi N$ appears in (\ref{Clausius}) can be interpreted as an 
interaction among the vortices~\cite{ManSMV}: each vortex effectively occupies an area of $4\pi$ (consistently with (\ref{Bradlow})), hence $N$ coexisting vortices have an area available for their motion which is a reduction of the area $A$ of the sphere by $N \times 4\pi$.
The virial coefficients associated to (\ref{Clausius})
are found to be all constant and equal to powers of $4\pi$:
\begin{equation}\label{virial}
PA=NT\sum_{\ell=0}^{\infty}(4\pi)^{\ell}n^{\ell};
\end{equation}
this virial expansion is 
reminiscent of the one for a gas of hard particles of finite size in a
one-dimensional box~\cite{ManSut}. Thus one might be tempted to think of 
the vortices in effective terms as rigid discs of area $4 \pi$ moving on the surface. But this picture already fails at first order in the expansion (\ref{virial}): the gas of hard disks would have first virial coefficient 
$8 \pi$~\cite{ReeHoo},
which is twice the coefficient of $n$ in the power series in (\ref{virial}).
In fact, a crucial difference between the gas of vortices and a gas of hard discs is that the shapes of the regions where the vortex density is concentrated become very different in the two cases whenever two or more particles come close together, and we expect the equation of state to be extremely sensitive to this.

Now we shall consider the statistical mechanics of the interacting regime
($\mu^2 \ne 0$). From our partition function (\ref{Zresult}), we can compute the thermodynamic average value of 
the observable $J \in C^{\infty}({\cal M}_N)$ given by (\ref{J}):
\begin{eqnarray}
\langle J \rangle & = & \frac{1}{Z}
\int_{T^{*}{\cal M}_N}J\;e^{-H/T}\,\frac{\omega_{\rm can}^{2N}}{(2N)!} \nonumber \\
&=&-T\frac{\partial}{\partial \mu^2} \log Z \nonumber \\
&=&-NT\left( \frac{A-4\pi N}{2T} \coth \frac{\mu^2 (A-4\pi N)}{2T} -\frac{1}{\mu^2}\right).  \label{expectJ}
\end{eqnarray}
Recall that $J$ is related to the height function on the sphere 
$x_{3}:=R\frac{1-|z|^2}{1+|z|^2} \in C^{\infty}(S^{2}_{R})$ by 
$J={\cal R}(R x_3)$ (cf.~(\ref{reduction})). Thus we propose to interpret the quantity ${J}/({2\pi N R})$ as an observable giving the height on $S^{2}_R$ of
configurations of $N$ vortices, and we shall denote it by $\tilde{x}_3$.
We find
\begin{equation} \label{expectx3}
\langle \tilde{x}_3 \rangle = \frac{\langle J \rangle}{2 \pi N R} =
-\left( 1-\frac{N}{R^2} \right)R \left( \coth \chi - \frac{1}{\chi}\right),
\end{equation}
where
\begin{equation} \label{chi}
\chi:= \frac{\mu^2 (A-4 \pi N)}{2 T}.
\end{equation}
The dependence of the average height on the parameter $\chi$ is plotted in Figure~1.

\begin{figure} \label{figheight}
\begin{center}
\vspace{1cm}
\includegraphics{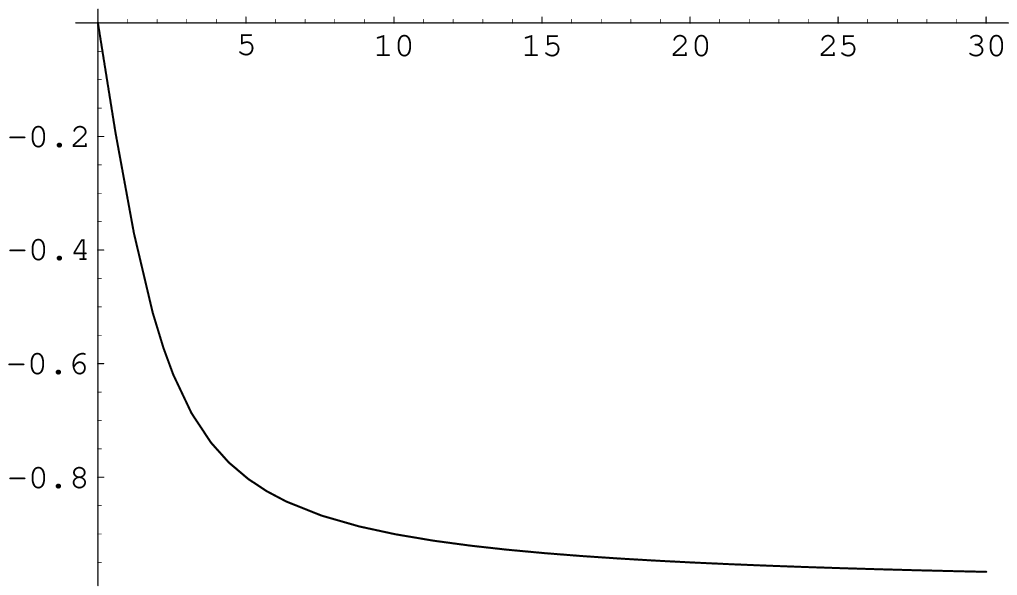}\\
\vspace{-6.9cm}\hspace{-10.6cm}$\langle \tilde{x}_3 \rangle / (1-\frac{N}{R^2})R$\\
\vspace{0cm}\hspace{11cm}$\chi$\\
\vspace{6.2cm}
\caption{\small \em 
The thermodynamic average $\langle \tilde{x}_3 \rangle$ 
of the height observable 
in the gas of $N$ vortices as a function of the parameter $\chi$.
}
\end{center}
\end{figure}

To interpret the meaning of (\ref{expectx3}), consider a simple model where
the vortex density is supported and homogeneously distributed on a spherical disc of area $4 \pi N$ and  centred at the minimum $z=\infty$ of the potential (\ref{Rx3}).
The height $x_3^{\rm max}$ of the points at the boundary of this disc is computed as
\[
2 \pi R \int_{-R}^{x_3^{\rm max}}dx_3=4\pi N \Longrightarrow x_{3}^{\rm max}=-R+\frac{2N}{R},
\]
where we made use of Archimedes' hat-box theorem.
In this model, the average height for the vortex density is then given by
\begin{equation}\label{calotte}
\langle x_3 \rangle^{\rm model}:=\frac{\int_{-R}^{x_3^{\rm max}}x_3 \,dx_3}
{\int_{-R}^{x_3^{\rm max}}dx_3}=
-\left( 1-\frac{N}{R^2} \right)R.
\end{equation}
Thus we find that
\[
\lim_{\chi \rightarrow \infty} \langle \tilde{x}_3 \rangle = \langle x_3 \rangle^{\rm model}. 
\]
In other words, the effective shape of an $N$-vortex distorts to a spherical
disc (of area  $4 \pi N$ and centred at the minimum of the potential $z=\infty$) as $\chi$ becomes large. This limit is typically achieved if the area $A-4\pi N$ available for the dynamics is large, the coupling $\mu^2$ is large, or the absolute temperature $T$ is low --- cf.~(\ref{chi}). Notice that in the picture where each vortex is approximated by a rigid disc of area $4 \pi$, the average height of the density distribution would be higher than 
(\ref{calotte}). The picture of the $N$-vortex as a disc of constant density
(localised at the bottom of the potential) provides a good description of the ground state of the system.

To describe a circularly symmetric
distribution of vortices on the sphere,
it is natural to introduce a vortex density function $\rho: [-R,R] \rightarrow 
\mathbb{R}$, defined on the interval of heights on $S^2_R$
and normalised as $\int_{-R}^{R}\rho \, dx_3=1$.
This function has parameters $A$, $T$ and $N$, and it could in principle be reconstructed as a Fourier--Legendre series from the partition function (\ref{Zresult}) if we could compute the reduction to the moduli space of all 
the powers of the height, ${\cal R}(f^m)$ for $m\in \mathbb{N}$.
We can write
\[
\int_{-R}^{R}x_3\,\rho(x_3)\,dx_3 :=
\langle \tilde{x}_3 \rangle =
\frac{\langle J \rangle}{2 \pi N R},
\]
which gives just the first Fourier--Legendre coefficient. The difference in
pressure $\Delta P$ between the highest and the lowest points of the gas on 
the sphere depends only on the trivial zeroth order coefficient: using the relation
\[
\nabla P +  N  \rho \, \nabla f =0
\]
for the pressure of a fluid of $N$ particles subject to a potential $f$
at equilibrium (this is analogous to the problem of fluid motion in a 
constant gravitational field, cf.~\cite{LifPitSP1} \S 25), we find
\begin{eqnarray*}
\Delta P & = & \int_{-R}^{R} \nabla P \, dx_3 =
-\int_{-R}^{R} N \rho(x_3) \frac{\partial}{\partial x_3}(\mu^2 R x_3) \, dx_3  = -\mu^2 NR <0.
\end{eqnarray*}

\section{Discussion} \label{secdisc}

We have been able to calculate the partition function of a gas of 
critically coupled abelian Higgs vortices interacting with an axially 
symmetric background potential on a sphere. When we switch off the interaction,
we recover Manton's partition function that gives physical insight on the
Liouville volume of vortex moduli spaces ${\cal M}_N$. Our study yields, in addition to Manton's formula (\ref{VolMN}) for ${\rm Vol}({\cal M}_N)$, an infinite number of nontrivial integrals (\ref{nontrivialints}) over 
${\cal M}_N$. These are additional data about the geometry of the moduli 
spaces, and they are all encapsulated by our partition function.
As an application, we computed the thermodynamic average height of the gas of $N$ vortices, and the result we found is consistent with the effective picture of the ground state as an $N$-vortex localised at the bottom of the potential 
as a spherical disc of constant density and area $4 \pi N$.

Our analysis was essentially an application of the Duistermaat--Heckman
localisation formula for a natural circle action on $({\cal M}_N,\omega)$,
in which the symplectic structure of the moduli space is a crucial ingredient.
There is an alternative model~\cite{Manfov} for dynamics of Ginzburg--Landau
vortices with a Schr\"odinger--Chern--Simons kinetic term, for which
${\cal M}_{N}$ (not $T^{\ast}{\cal M}_N$) plays the r\^ole of phase space
in the adiabatic approximation;
in this context, the K\"ahler form $\omega$ appears naturally as a symplectic structure~\cite{RomQCS}. Our work illustrates that the symplectic point of view can also be fruitful in the study of the abelian Higgs model.

We have already noted that our formula (\ref{nontrivialints}) can be 
applied to calculate integrals of
general circularly symmetric functions on the moduli space. 
One may therefore hope to study the interaction of the
vortices with any symmetric potential $f$ on $S^{2}_{R}$ --- or perhaps
even an ${\rm SO}(3)$-invariant intervortex interaction modelling the 
Ginzburg--Landau potential at 
$\lambda^2\ne 1$ to some degree of approximation, which would be an 
obviously interesting extension of our work~\cite{EasRomCFHM}.
Analytical results about the Ginzburg--Landau potential (for arbitrary $N$) are already available \cite{SpeSIF,RomSpeSSD}, but they refer to the situation where the
vortices are well separated on the plane. A treatment of the interaction to include the interesting effects near clustering configurations on the moduli spaces will almost certainly need to use some
numerical input~\cite{ShaSVN}. It is believed that abelian Higgs vortices even slightly away from critical coupling should satisfy
a realistic equation of state such as the van der Waals equation, which
accounts for phase transitions. Progress in this direction would shed light 
on the phenomenology of thin superconductors.

\vspace{.75cm}

\noindent
{\Large \bf Acknowledgements}\\[4mm]
\noindent
The author is thankful to Nick Buchdahl, Nick Manton and Michael Murray 
for useful comments. 
This work was supported by the Australian Research Council.

\vspace{.75cm}

\begin{small}

\bibliographystyle{numsty}
\bibliography{biblio}

\end{small}

\end{document}